\documentclass[conference]{IEEEtran}
\IEEEoverridecommandlockouts
\usepackage{cite}
\usepackage{url}
\usepackage{balance}
\usepackage{booktabs}
\usepackage{amsmath, amssymb, amsfonts}
\usepackage{algorithmic}
\usepackage{algorithm}
\usepackage{verbatim}
\usepackage{multirow}
\usepackage{bbding}
\usepackage{bm}
\usepackage{graphicx}
\usepackage{textcomp}
\usepackage{xcolor}
\usepackage{pifont}
\usepackage{multirow}
\usepackage{footmisc}

\def\BibTeX{{\rm B\kern-.05em{\sc i\kern-.025em b}\kern-.08em T\kern-.1667em\lower.7ex\hbox{E}\kern-.125emX}}

\begin{document}
    
    \title{LP-Spec: Leveraging LPDDR PIM for Efficient LLM Mobile Speculative Inference with Architecture-Dataflow Co-Optimization}
    
    \renewcommand{\thefootnote}{\fnsymbol{footnote}}
    
    \author{
        \IEEEauthorblockN{
            Siyuan He\textsuperscript{*}, 
            Zhantong Zhu\textsuperscript{*}, 
            Yandong He, 
            Tianyu Jia\textsuperscript{\dag}}
        \IEEEauthorblockA{\textit{School of Integrated Circuits} \\
        \textit{Peking University}\\
        Beijing, China\\
        \textsuperscript{\dag}Corresponding Email: tianyuj@pku.edu.cn}
    }

    \vspace{-20pt}

    \maketitle
    \footnotetext[1]{Equal contribution to this work.}
    
    \begin{abstract}
        LLM inference on mobile devices faces extraneous challenges due to limited memory bandwidth and computational resources.
        To address these issues, speculative inference and processing-in-memory (PIM) techniques have been explored at the algorithmic and hardware levels. However, speculative inference results in more compute-intensive GEMM operations, creating new design trade-offs for existing GEMV-accelerated PIM architectures.
        Furthermore, there exists a significant amount of redundant draft tokens in tree-based speculative inference, necessitating efficient token management schemes to minimize energy consumption.
        In this work, we present \textit{LP-Spec}, an architecture-dataflow co-design leveraging hybrid LPDDR5 performance-enhanced PIM architecture with draft token pruning and dynamic workload scheduling to accelerate LLM speculative inference.
        A near-data memory controller is proposed to enable data reallocation between DRAM and PIM banks. Furthermore, a data allocation unit based on the hardware-aware draft token pruner is developed to minimize energy consumption and fully exploit parallel execution opportunities.
        Compared to end-to-end LLM inference on other mobile solutions such as mobile NPUs or GEMV-accelerated PIMs, our \textit{LP-Spec} achieves 13.21$\times$, 7.56$\times$, and 99.87$\times$ improvements in performance, energy efficiency, and energy-delay-product (EDP). Compared with prior AttAcc PIM and RTX 3090 GPU, \textit{LP-Spec} can obtain 12.83$\times$ and 415.31$\times$ EDP reduction benefits.
    \end{abstract}


    \section{Introduction}

Large language models (LLMs) have rapidly advanced across multiple application domains and exhibited remarkable performance \cite{brown_language_2020, chatpgt, github_copilot}. Recently, LLMs have transitioned from large-scale cloud infrastructures to consumer-level mobile devices, such as smartphones or laptops \cite{apple_intelligence, samsung_live_translate}. Mobile devices feature dedicated AI acceleration hardware such as neural processing units (NPUs) to handle computation demands of AI tasks, including LLM inference. However, mobile hardware must adhere to stringent energy, area, and inference latency constraints. With limited memory bandwidth, memory and computation capacity, efficient LLM deployment on mobile devices necessitates cross-stack optimization approaches.

\begin{figure}[t]
  \centering
  \includegraphics[width=\linewidth]{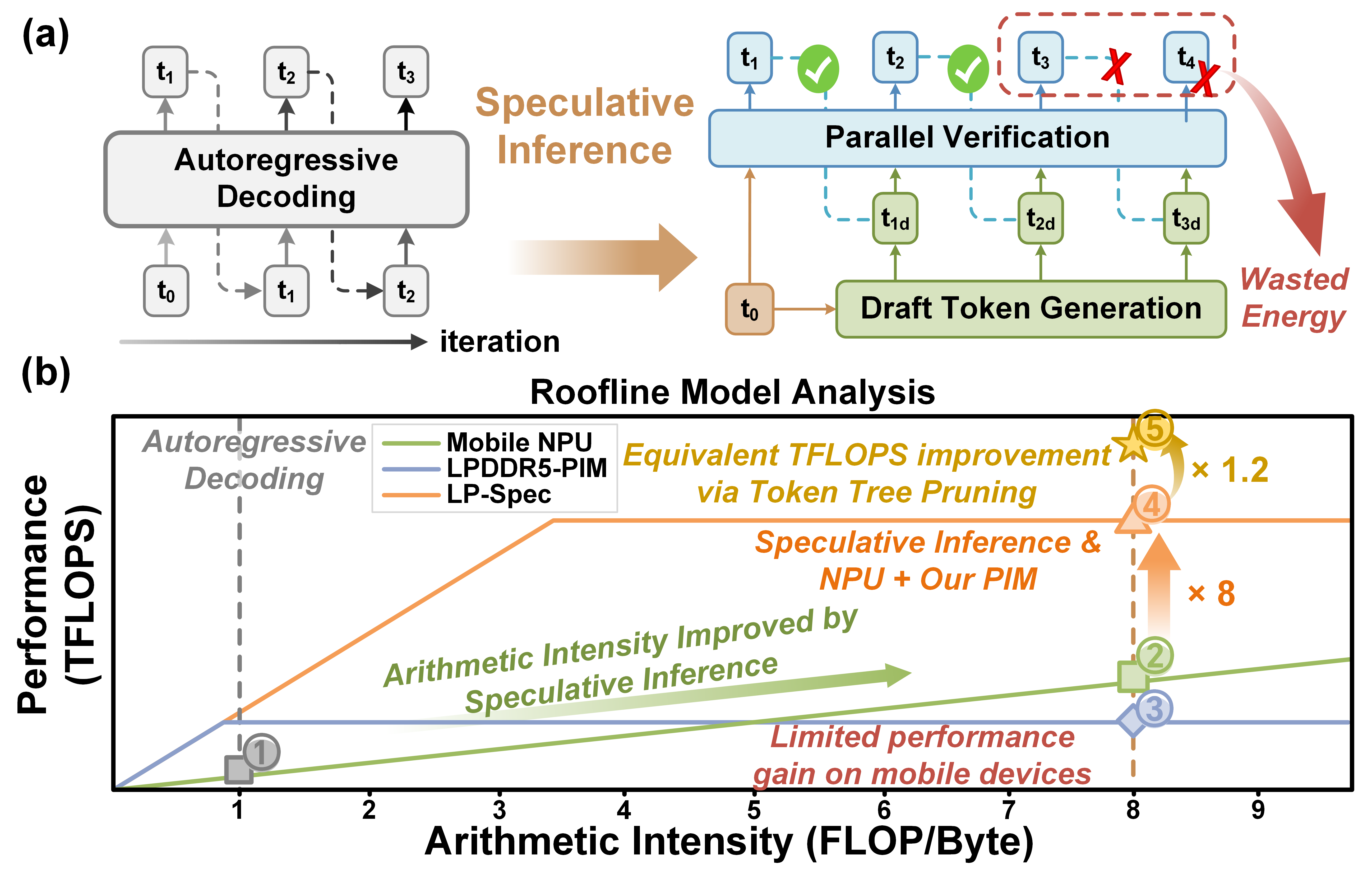}
  \vspace{-10pt}
  \caption{(a) Autoregressive decoding and speculative inference, (b) challenges of speculative inference on existing mobile devices and our design target.}
  \label{fig: mobile_LLM_design_challenge}
  \vspace{-10pt}
\end{figure}

To enable efficient LLM inference on mobile devices, significant algorithm-level advances have emerged. Beyond established techniques such as quantization \cite{dettmers_llmint8_2022, dettmers_qlora_2023, xiao_smoothquant_2024} and weight pruning \cite{ma_llm-pruner_2023, frantar_sparsegpt_2023, sun_simple_2024}, speculative inference \cite{miao_specinfer_2024, zhong_propd_2024, leviathan_fast_2023, chen_accelerating_2023, cai_medusa_2024, xia_speculative_2023} has recently emerged as a promising approach to overcome the sequential bottleneck of autoregressive decoding. As shown in Fig. \ref{fig: mobile_LLM_design_challenge}(a), conventional autoregressive decoding generates one token at a time, with each step depending on the previous output. Speculative inference, by contrast, predicts multiple future tokens and verifies them in parallel, achieving around 2-3$\times$ speedup over vanilla autoregressive decoding. 

Fig. \ref{fig: mobile_LLM_design_challenge}(b) utilize a roofline model to illustrate LLM inference deployment on mobile devices. Conventional autoregressive decoding is the primary contributor to LLM inference latency \cite{yuan_llm_2024} due to limited memory bandwidth (Fig. \ref{fig: mobile_LLM_design_challenge}(b) \ding{182}).
Speculative inference addresses this by transforming general matrix-vector multiplication (GEMV) operations into more compute-intensive general matrix-matrix multiplication (GEMM) operations (Fig. \ref{fig: mobile_LLM_design_challenge}(b) \ding{183}). 
However, resource and power-constrained mobile devices typically only incorporate LPDDR5 DRAM modules with limited bandwidth, e.g., 51.2 GB/s, and rely on mobile NPUs for LLM inference, resulting in sub-optimal performance gains (\ding{183}).
Furthermore, increasing the number of draft tokens, i.e., speculation length, also raises the computational burden for verification, and many tokens are ultimately rejected, leading to wasted energy. 

To address this memory bandwidth bottleneck, recent advances in DRAM-based processing-in-memory (PIM) technology have been introduced, leveraging high internal bandwidth by enabling computation directly inside DRAM dies.
However, current PIM designs face two key limitations for enabling efficient LLM speculative inference on mobile devices. \textit{First}, prevalent PIM architectures are predominantly optimized for GEMV operations, as seen in SK Hynix's GDDR6-PIM \cite{lee_1ynm_2022} and Samsung's HBM2-PIM \cite{kim_aquabolt-xl_2022}. This GEMV-centric architecture is inefficient for speculative inference, which involves a higher proportion of GEMM operations (\ding{184}). \textit{Second}, current PIM architectures prevent NPU and PIM parallel execution, as DRAM memory access is blocked by PIM operations, thereby limiting overall hardware utilization. Consequently, novel PIM architectural innovations and dynamic scheduling schemes are desired to support efficient deployment of speculative inference on mobile devices.

In this work, we introduce \textit{LP-Spec}, an architecture-dataflow co-design leveraging hybrid \underline{LP}DDR5-based PIM architecture with dynamic workload scheduling and runtime data reallocation to accelerate LLM \underline{spec}ulative inference.
First, we present a hybrid LPDDR5-PIM module with DRAM and PIM ranks and near-data memory controller. Then, we propose a hardware-aware draft token pruning scheme to enhance computational efficiency and inference latency.
Given varying speculation lengths at runtime, we introduce a dynamic workload scheduler with data reallocation to fully exploit NPU-PIM parallel execution and maximize hardware utilization.
We evaluate \textit{LP-Spec} against prior NPU and NPU-PIM systems \cite{park_multi-mode_2022, huang_5g_2023, kim_aquabolt-xl_2022, kim_samsung_2023}, showing up to 13.21$\times$ performance improvement and 7.56$\times$ better energy efficiency. Compared with prior PIM-based design AttAcc \cite{park_attacc_2024} and Nvidia RTX 3090 GPU, \textit{LP-Spec} achieves 12.83$\times$ and 415.31$\times$ EDP reduction.

The contributions of our work are summarized as follows:
\begin{itemize}

  \item We present \textit{LP-Spec}, an NPU-PIM heterogeneous architecture with GEMM-enhanced LPDDR5-PIM to improve computational performance for speculative inference.

  \item We design a near-data memory controller to support simultaneous PIM computation and DRAM access, enabling data reallocation at runtime.

  \item We present a hardware-aware runtime draft token pruning scheme to minimize redundant computation and energy consumption cost.

  \item We evaluate our design against mobile NPU and PIM baselines, showing up to 99.87$\times$ EDP gain. Our \textit{LP-Spec} also shows 12.83$\times$ and 415.31$\times$ better EDP than prior AttAcc PIM and RTX 3090 GPU.
\end{itemize}

\section{Background}

\subsection{Processing-in-Memory (PIM)}

Conventional architectures like GPUs face performance bottlenecks for memory-intensive operations due to restricted memory bandwidth.
PIM designs address this challenge by integrating processing units within or near DRAM banks \cite{mcdram1, mcdram2, lee_1ynm_2022, kim_aquabolt-xl_2022}.
By leveraging high internal bandwidth of DRAM banks and reducing off-chip data movement, PIM significantly improves computation throughput and energy efficiency.
For instance, while a single $\times$64 LPDDR5 DRAM chip with 4 $\times$16 LPDDR5 DRAM dies provides 51.2 GB/s of external I/O bandwidth, its internal all-bank bandwidth can reach up to 409.6 GB/s \cite{lpddr5_spec}.
Moreover, data transfers within DRAM consume only 15\% of the energy required for off-DRAM transfers \cite{kim_samsung_2023}.
Recent PIM-based system architectures have explored combining PIM with xPUs, e.g., GPUs and NPUs, to optimize LLM inference \cite{park_attacc_2024, heo_neupims_2024, li_specpim_2024}. For example, AttAcc \cite{park_attacc_2024} integrates GEMV and Softmax units within HBM stacks to accelerate attention layers, employing optimized pipelining and parallelism strategies to improve autoregressive decoding. Similarly, SpecPIM \cite{li_specpim_2024} explores the architectural and workload mapping design space to enhance draft language model-based (DLM) speculative inference.

Prior work focused on leveraging GPU-PIM heterogeneous systems to accelerate large-scale LLM inference for cloud serving scenarios, in which the compute-intensive FC layers are executed in GPUs and the memory-intensive attention layers are offloaded to PIM.
However, these approaches are less effective when adpated to mobile devices, due to the much lower memory bandwidth and limited computational resources. 
Mobile NPUs are often limited by off-chip memory bandwidth while conventional GEMV-targeted PIMs struggle with arithmetic-intensive GEMM operations.
This limitation motivates us to explore architecture-dataflow co-optimization for speculative LLM inference on resource-constrained mobile devices.

\subsection{Speculative Inference}

\begin{table}[t]
  \caption{\textbf{Comparison between different types of speculative inference.}}
  \label{tab: speculative_comparison}
  \centering
  \resizebox{0.85\linewidth}{!}{
  \begin{tabular}{cccc}
    \toprule
    \textbf{Methods} & \textbf{\begin{tabular}[c]{@{}c@{}}Drafting \\ Scheme\end{tabular}} & \textbf{\begin{tabular}[c]{@{}c@{}}Token \\ Structure\end{tabular}} \\
    \midrule
    SpecDec \cite{xia_speculative_2023} & DLM & Sequence-based \\ 
    Blockwise \cite{stern_blockwise_2018} & Decode Heads & Sequence-based \\
    SpecInfer \cite{miao_specinfer_2024} & DLM & Tree-based \\ 
    Medusa \cite{cai_medusa_2024} & Decode Heads & Tree-based \\
    \bottomrule
  \end{tabular}
  }
\end{table}

LLM inference typically employs an autoregressive decoding scheme, generating one output token per iteration due to the data dependencies between adjacent tokens. This sequential nature makes the inference predominantly memory-bound, as each forward pass requires accessing the complete model parameters as well as Key-Value caches (KV caches). To address the inefficiencies of autoregressive decoding, \textit{Draft-then-Verify} scheme, namely speculative inference, has been introduced \cite{miao_specinfer_2024, zhong_propd_2024, leviathan_fast_2023, chen_accelerating_2023}.
In speculative inference, multiple future tokens are first \textit{drafted}, followed by parallel \textit{verification} against the target language model (TLM).

\begin{figure}[t]
  \centering
  \includegraphics[width=\linewidth]{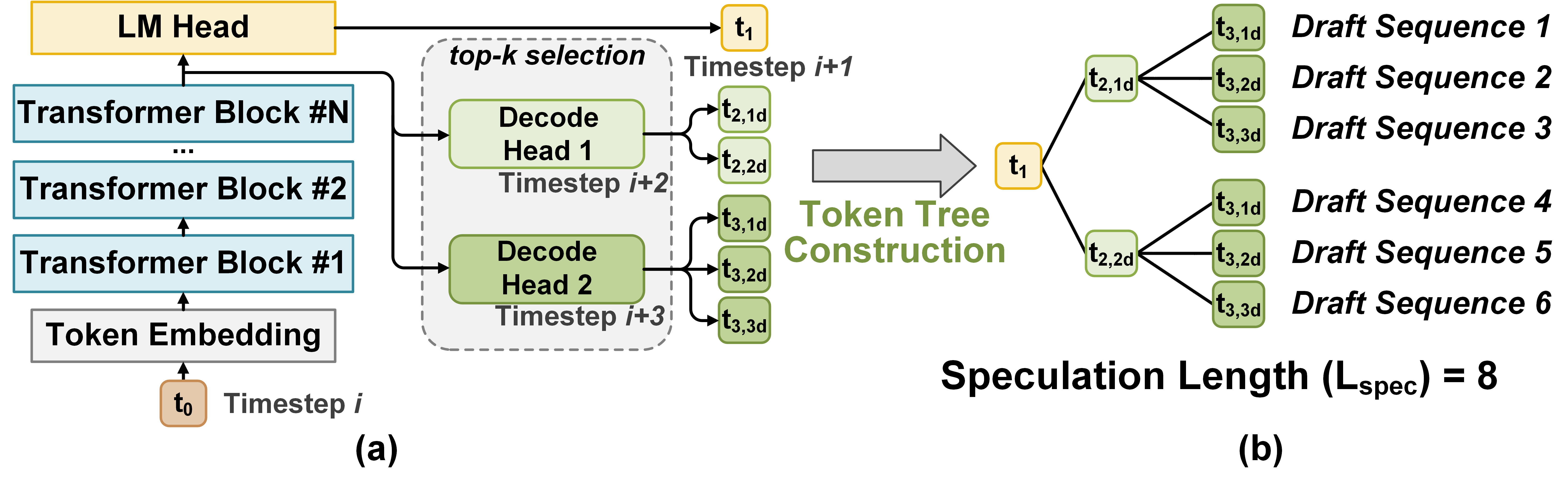}
  \vspace{-15pt}
  \caption{Tree-based speculative inference and token tree structure.}
  \vspace{-15pt}
  \label{fig: tree-based speculative inference}
\end{figure}

Speculative inference has been widely studied using various drafting schemes and token structures, as listed in Table \ref{tab: speculative_comparison}. For drafting schemes, one approach leverages a small DLM to generate draft tokens over several iterations \cite{leviathan_fast_2023, chen_accelerating_2023, li_specpim_2024}, verified in parallel by the original TLM,
Alternatively, \textit{self-drafting} \cite{stern_blockwise_2018, cai_medusa_2024, zhong_propd_2024} augment the TLM itself by integrating additional Feed-Forward Network (FFN) heads (henceforth referred to as Decode Heads) on the final Transformer block, as shown in Fig. \ref{fig: tree-based speculative inference}(a).

For token structures, a conventional sequence-based token structure leads to limited acceptance rate. SpecInfer \cite{miao_specinfer_2024} introduced a tree-based token structure to improve the acceptance length of draft tokens, which selects Top-\textit{k} draft tokens from each Decode Head and merges them into a token tree by sharing common prefixes. As illustrated in Fig. \ref{fig: tree-based speculative inference}(b), at timestep \textit{i}, the original LM head generates one output token for timestep \textit{i+1}, while Decode Head 1 and 2 are used to predict future tokens at timestep \textit{i+2} and \textit{i+3}, respectively. Consequently, Top-2 predictions of Decode Head 1 and Top-3 predictions of Decode Head 2 result in 2$\times$3 draft sequences represented as a token tree. The draft tokens are verified in parallel, where a portion of draft tokens are accepted. The parallel verification increases the arithmetic intensity, as the model parameters and KV-caches are accessed only once for multiple draft tokens, which helps to alleviate the memory bandwidth bottleneck in mobile LLM inference.

    \section{Motivation}

\subsection{Analysis of Speculative Inference on PIM Designs}
PIM designs leverage high bank-level parallelism to deliver significant throughput advantages for GEMV operations compared to conventional architectures like GPUs.
However, speculative inference increases the number of parallel tokens processed during the decoding stage, effectively converting GEMV operations into GEMM operations, degrading the efficiency of prior PIM designs.
To quantify this effect, we profile the performance of Samsung LPDDR5-PIM \cite{kim_aquabolt-xl_2022} and a mobile NPU for Llama2-7B inference using AttAcc-like data mapping \cite{park_attacc_2024}.
Detailed configurations are shown in Table \ref{tab: hardware specification}.

\begin{figure}[t]
    \centering
    \includegraphics[width=0.95\linewidth]{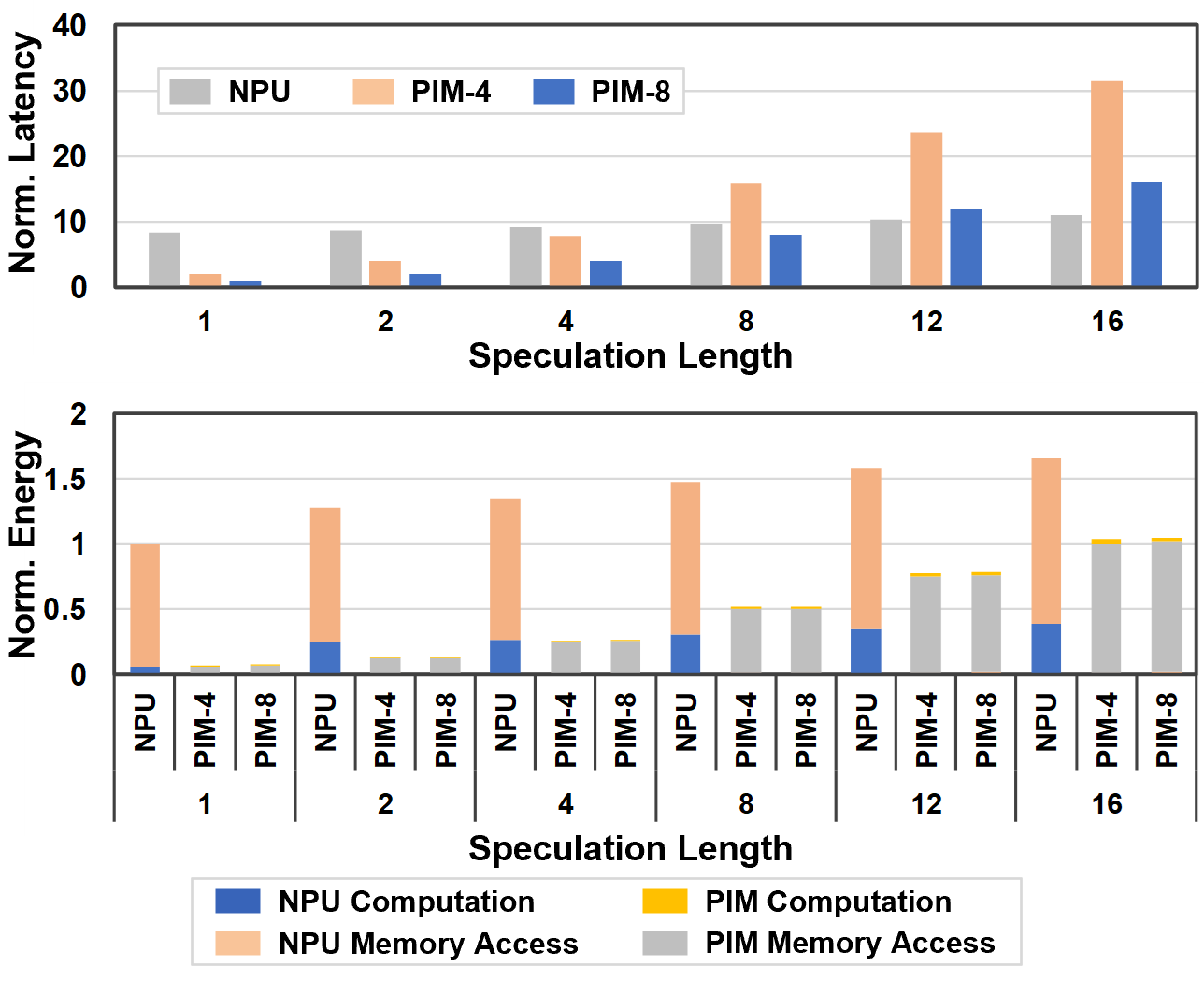}
    % \vspace{-5pt}
    \caption{Speculative inference latency and energy breakdown.}
    % \vspace{-10pt}
    \label{fig: pim profile}
\end{figure}

Fig. \ref{fig: pim profile} illustrates the performance impact of varying speculation lengths, i.e., the number of speculative tokens, evaluated across 4-die (PIM-4) and 8-die (PIM-8) configurations.
For one decoding iteration, PIM-4 achieves 4.25$\times$ latency improvement and 15.4$\times$ energy reduction over mobile NPUs, while PIM-8 achieves 8.34$\times$ latency and 15.2$\times$ energy gains.
However, as the speculation length increases from 1 to 16, both latency and energy advantages deteriorate sharply across PIM configurations.

This performance degradation stems from increased arithmetic intensity of GEMM operations, which improves NPU utilization while eroding bandwidth advantages of PIM designs.
Current PIM designs are primarily optimized for GEMV operations, relying on per-bank vector units with limited computational throughput. Their inability to exploit data reuse in GEMM workloads ultimately diminishes effective bandwidth gains for LLM speculative inference.

Although naively increasing the number of PIM devices can also boost overall performance, this approach introduces significant cost and capacity trade-offs. The compute units in PIM devices are fabricated using low-density DRAM process technology, which reduces DRAM memory capacity.
For example, Samsung HBM-PIM devices reduce per-die capacity from 8 Gb to 4 Gb \cite{kim_aquabolt-xl_2022}. To achieve cost-efficient deployment of speculative inference on mobile devices, it is critical to fully exploit the hardware resources, i.e., both NPU and PIM, through effective workload scheduling. Additionally, architectural optimizations for PIM devices are essential to enhance the performance of GEMM operations.

\subsection{Dynamic Token Pruning and Scheduling Opportunities}

\begin{figure}[t]
    \centering
    \includegraphics[width=\linewidth]{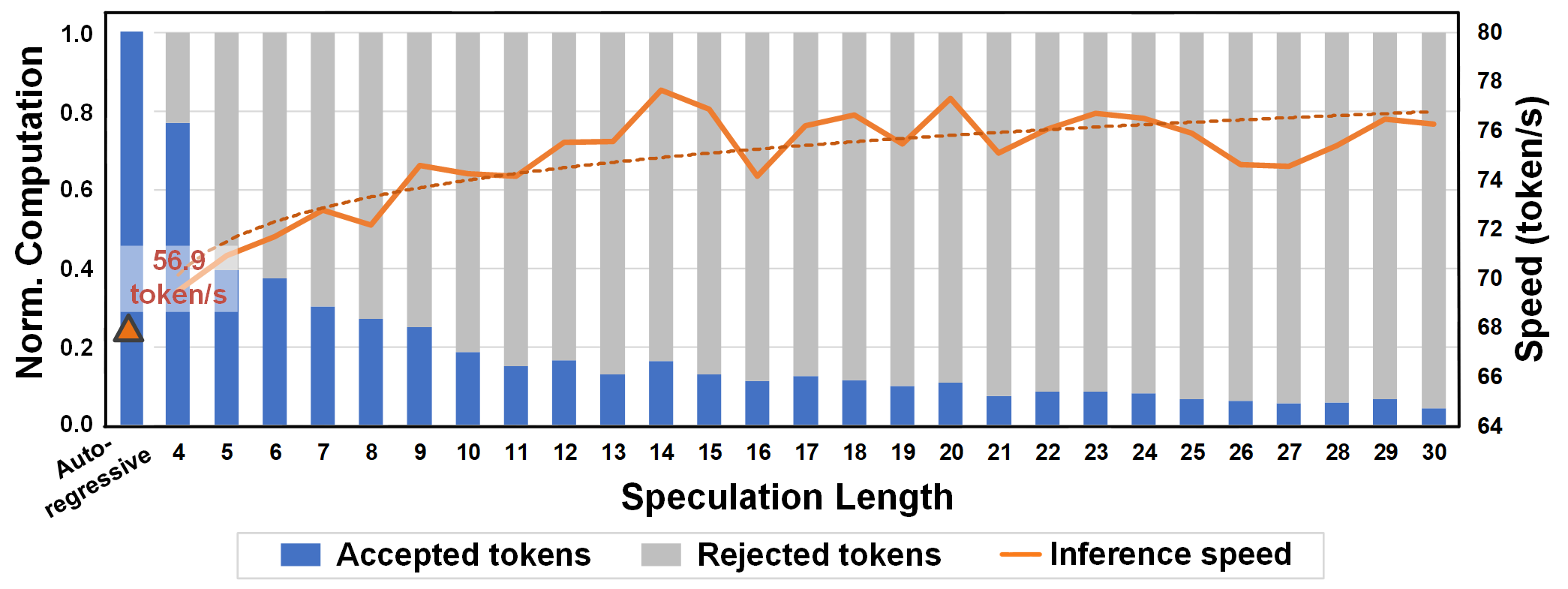}
    \vspace{-10pt}
    \caption{Profiling of tree-based speculative inference.}
    \vspace{-10pt}
    \label{fig: medusa profiling}
\end{figure}

\begin{figure*}[t]
  \centering
  \includegraphics[width=1\linewidth]{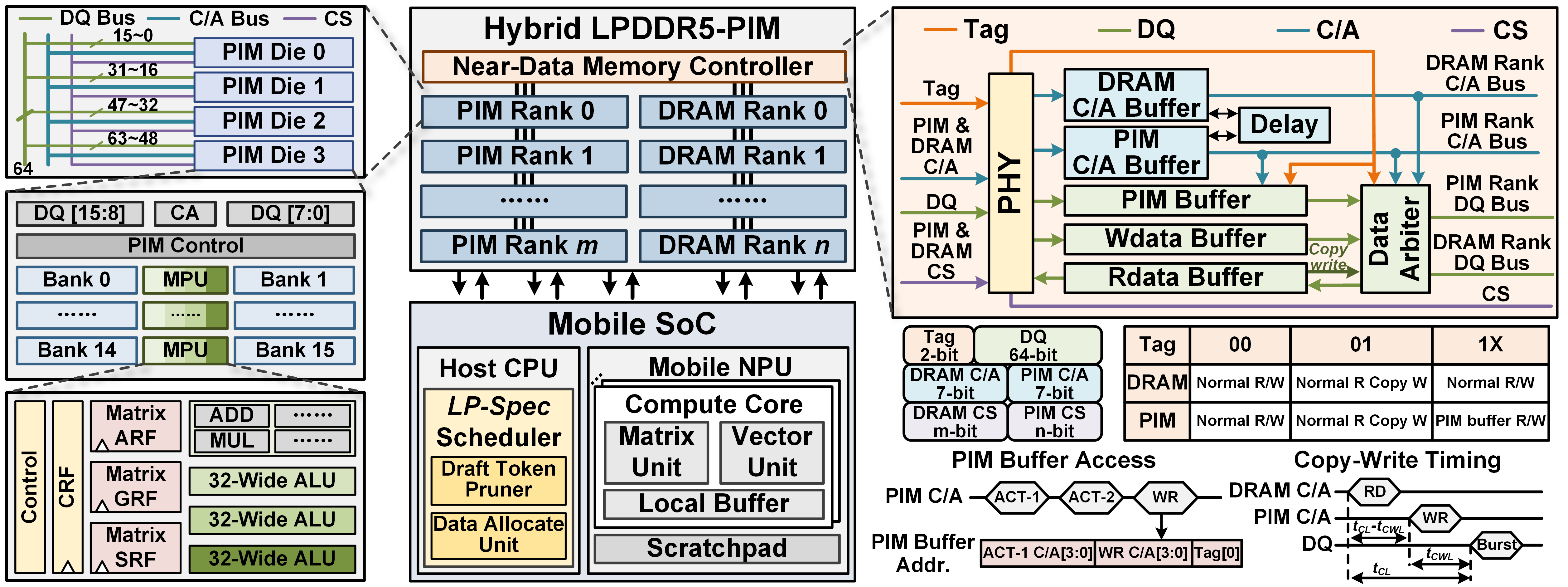}
  % \vspace{-10pt}
  \caption{
  Overview of \textit{LP-Spec} architecture, the design details of the LP-spec PIM and the near data memory controller.}
  \label{fig: lp-spec overview}
  % \vspace{-10pt}
\end{figure*}

In tree-based speculative inference, the number of draft tokens grows exponentially with the depth of the token tree. To analyze its inference characteristics, we profile Medusa \cite{cai_medusa_2024} on an NVIDIA RTX 4090 using randomly generated dense tree structures. 
It is observed in Fig. \ref{fig: medusa profiling} that expanding the token tree will increase inference speedup.
However, the portion of computation for verifying rejected tokens is also growing. Since rejected tokens are not reused in subsequent decoding iterations, this results in computational and energy waste.

These observations highlight the need for more effective token optimization strategies. Prior approaches have explored techniques such as predetermined optimized token tree structures \cite{cai_medusa_2024} or using early token pruning mechanisms \cite{zhong_propd_2024}. However, these methods either fail to capitalize on dynamic optimization opportunities or neglect hardware constraints of PIM-based mobile devices.
To address this gap, we explore a hardware-aware draft token pruner which adaptively optimizes the token tree structure to balance inference speed and energy efficiency based on runtime conditions. 

Based on the above analysis, we can observe that speculative inference on PIM-based mobile devices remains challenging due to limited computational performance of PIM designs and the redundancies of speculative tokens.
In this work, we propose \textit{LP-Spec}, an architecture-dataflow co-optimization approach for efficient speculative inference. 
At the hardware level, we introduce architectural enhancements including bank-level matrix processing units and a near-data memory controller enabling concurrent PIM and NPU execution.
At the software level, we propose dynamic token pruning and scheduling strategies to improve both the performance and hardware utilization of speculative inference on PIM-based mobile platforms.

    \section{LP-Spec Architecture Design}

\subsection{Architecture Overview}

In this work, we present \textit{LP-Spec}, an architecture-dataflow co-design aimed at facilitating efficient speculative inference for LLMs on mobile platforms.
As shown in Fig. \ref{fig: lp-spec overview}, \textit{LP-Spec} integrates a mobile SoC with a hybrid LPDDR5-PIM module. The SoC includes both a mobile NPU and a host CPU, with communication between the SoC and PIM facilitated via a 64-bit data bus. The NPU architecture refers to mobile NPUs \cite{park_multi-mode_2022, huang_5g_2023} and features a last-level scratchpad memory and 16 compute cores, each equipped with a SIMD-style matrix unit, a vector unit, and a local buffer.
To mitigate the impact of PIM devices on DRAM capacity, the hybrid LPDDR5-PIM module incorporates multiple DRAM and PIM ranks for a total capacity of 16 GB, thus balancing cost and performance.
This memory capacity enables the deployment of edge LLMs such as Llama2-7B \cite{touvron_2023_llama2} at INT8 data precision.
Each DRAM or PIM rank consists of four dies connected in parallel, operating in lockstep with the same command and address (C/A) and chip select (CS) signals, while utilizing distinct data lines, following JEDEC LPDDR5 specifications \cite{lpddr5_spec}.
Additionally, the hybrid LPDDR5-PIM module also features a near-data memory controller (NMC), enabling dynamic data reallocation between DRAM and PIM ranks at low cost.

The host incorporates the \textit{LP-Spec} scheduler, which consists of a draft token pruner (DTP) and data allocation unit (DAU). The DTP generates an optimized token tree after each decoding iteration, resulting in varying speculation lengths. The DAU dynamically reallocates model weights between DRAM and PIM ranks through the NMC, ensuring synchronized parallel execution of PIM and NPU devices.

\subsection{PIM Microarchitecture and Data Mapping}

Fig. \ref{fig: lp-spec overview} left details the design of an LPDDR5-PIM die.
Building on existing commodity LPDDR5-PIM microarchitecture and instruction set architecture (ISA) \cite{kim_aquabolt-xl_2022, lee_hardware_2021}, we extend and enhance PIM functionality to accelerate speculative inference while addressing the constraints of mobile devices.
Each PIM die integrates PIM control logic, 16 banks, and 8 matrix processing units (MPUs), with each MPU shared between two banks.
An MPU consists of three key components: (1) four 32-wide SIMD ALUs, (2) command, general, accumulate, scalar register files (CRF, GRF, ARF, and SRF), and a controller.
Given the prevalence of quantized LLM models for edge deployment, the 32-wide ALU features 32 INT8 multipliers and adders.
Since an INT8 MAC unit occupies only 26.5\% of the area and consumes 63.6\% of the energy of a FP16 MAC in 20 nm DRAM process\cite{lee_hardware_2021}, we integrate four 32-wide ALUs while maintaining acceptable hardware cost.
The register files consist of 32 32-bit CRFs, 16 4$\times$256-bit matrix GRFs, 16 4$\times$8-bit matrix SRFs, and 8 4$\times$1024-bit matrix ARFs. 
The ARFs extend precision from INT8 to INT32 for accumulation.
The controller fetches PIM instructions from the CRF and schedules the MPU accordingly.
Similar to Samsung’s LPDDR5-PIM \cite{kim_aquabolt-xl_2022}, the MPU supports source operands from both register files and DRAM banks, with bank-sourced data broadcast to all four ALUs for parallel computation. 
As most power consumption during PIM execution arises from DRAM access ($>$ 90\%) \cite{papi}, the MPU power consumption increases only by 23.2\% due to more DRAM data reuse compared with HBM-PIM \cite{li_specpim_2024, lee_hardware_2021}, which remains under the DRAM power budget.
When executing PIM computations, all PIM ranks are first switched into all bank mode to load data and instructions, after which they switch into all bank PIM mode to trigger the execution of PIM instructions. This mode switching is achieved by configuring the mode registers.

\begin{figure}[t]
  \centering
  \includegraphics[width=0.9\linewidth]{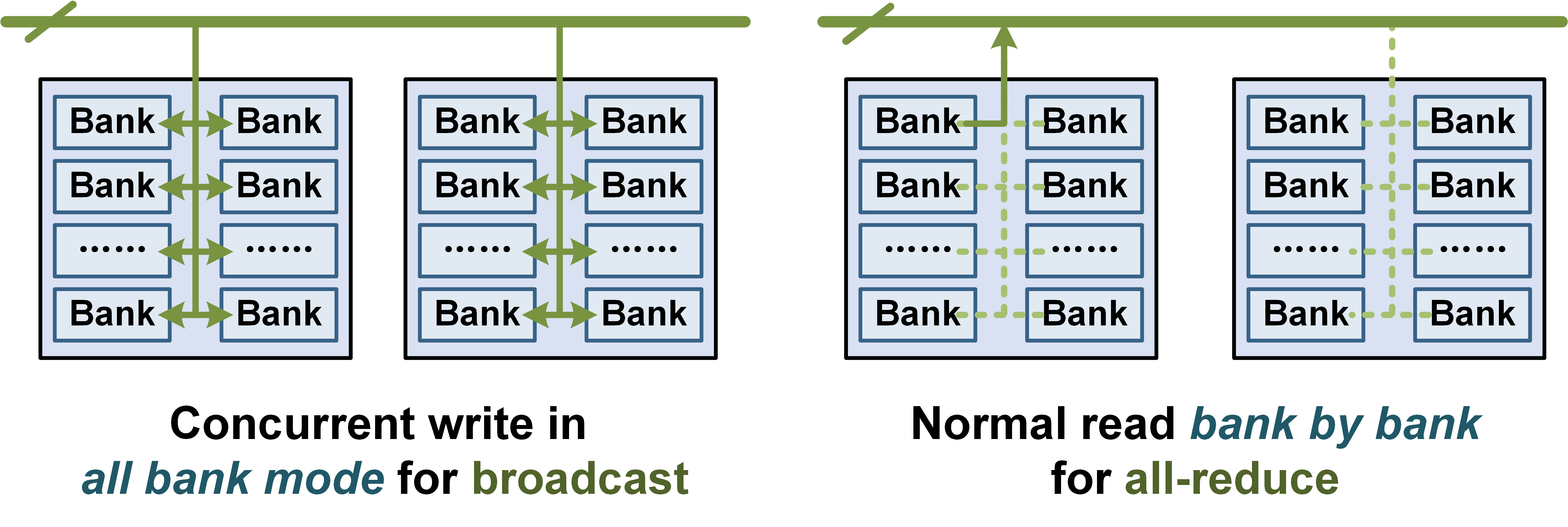}
  \vspace{-5pt}
  \caption{Comparison of communication overhead between broadcast and all-reduce operations in LPDDR5-PIM architecture.}
  \label{fig: data_mapping}
  \vspace{-10pt}
\end{figure}

Considering data mapping in PIM, column-wise partitioning minimizes output data transfer but requires broadcasting of the input data, while row-wise partitioning reduces the input data transfer but involves all-reduce of the output data.
Prior work AttAcc \cite{park_attacc_2024} employs both row-wise and column-wise partitions for K/V mapping in its customized HBM-PIM, utilizing the accumulators on the bank group and buffer die to perform all-reduce operations.
For the LPDDR5-PIM architecture, broadcast operations are easily implemented by enabling all bank mode and activating all CS signals, as shown in Fig. \ref{fig: data_mapping}.
Without on-die accumulators, the PIM execution units can only communicate with other banks on the same die or across different dies through the host.
Consequently, all-reduce operations require significantly more data transfer than broadcast operations, resulting in far lower effective bandwidth.
For instance, when using 8 PIM dies with 8 compute units per die, all-reduce operations incur 64$\times$ greater data transfer than broadcast operations.
Hence, we adopt column-wise partitioning to minimize communication costs.

\subsection{Near-Data Memory Controller}

Fig. \ref{fig: lp-spec overview} top right shows the architecture design of the NMC.
The NMC serves two key functions: enabling concurrent PIM computation and standard DRAM access for the SoC, as well as efficient data reallocation between PIM and DRAM ranks.
To support simultaneous PIM computation and DRAM access, the NMC receives independent C/A signals for DRAM ranks and PIM ranks, allowing for parallel operation.
Since the SoC can only access either the DRAM memory space or the PIM memory space at any given time, PIM and DRAM can share the same DQ lines.
However, larger input matrices in speculative inference may necessitate reloading both the input and the partial sums of the output, leading to contention for the data bus with the NPU.
To mitigate this problem, the NMC integrates a dedicated 4 KB PIM global buffer for PIM read/write operations.
Unused C/A pins are repurposed as addresses for this global buffer.
When execution times between the NPU and PIM become imbalanced, model parameters need to be reallocated between DRAM and PIM ranks.
A naive approach would involve reading the model parameters to host and then writing them back to DRAM or PIM ranks, which introduces additional latency and IO energy consumption.
In contrast, the NMC introduces a feed-forward path from the read data buffer to the write data arbiter, enabling in-situ copying of data read from DRAM/PIM to PIM/DRAM.

To support NMC functions, additional tag bits are required for guidance, and DRAM timing constraints must be carefully considered.
As shown in Fig. \ref{fig: lp-spec overview}, we add a 2-bit tag to indicate the operation mode: `00' for normal read/write, `01' for copy-write, and `1x' for PIM global buffer read/write.
For PIM global buffer operations, the LSB of the tag combines with an 8-bit address for bank and bank group in ACT-1, WR, and RD commands, forming a 9-bit address to access the PIM global buffer.
To ensure proper timing between RD and WR commands in copy-write operations, specific DRAM timing constraints must be adhered to.
The interval between issuing an RD command and the start of data transmission is defined as $t_{\textit{CL}}$, while $t_{\textit{CWL}}$ specifies the delay between a WR command and the beginning of the write data.
Thus, a delay of $t_{\textit{CL}}-t_{\textit{CWL}}$ should be inserted between the RD and WR.

    \section{LP-Spec Scheduler}
\label{sec: lp-spec scheduler}
Our observations indicate that substantial energy is wasted on verifying rejected tokens. To enhance runtime efficiency of LLM speculative inference, we propose a hardware-aware draft token pruner (DTP).
Based on dynamic speculation lengths, we design a data allocation unit (DAU) to facilitate synchronized NPU-PIM parallel execution.
Fig. \ref{fig: scheduler framework} illustrates the overall framework of the \textit{LP-Spec} scheduler.

\subsection{Draft Token Pruner}
The DTP employs a token tree accuracy model and runtime estimator to dynamically optimize the token tree through iterative sampling and pruning after each decoding step.

\begin{figure}[t]
  \centering
  \includegraphics[width=\linewidth]{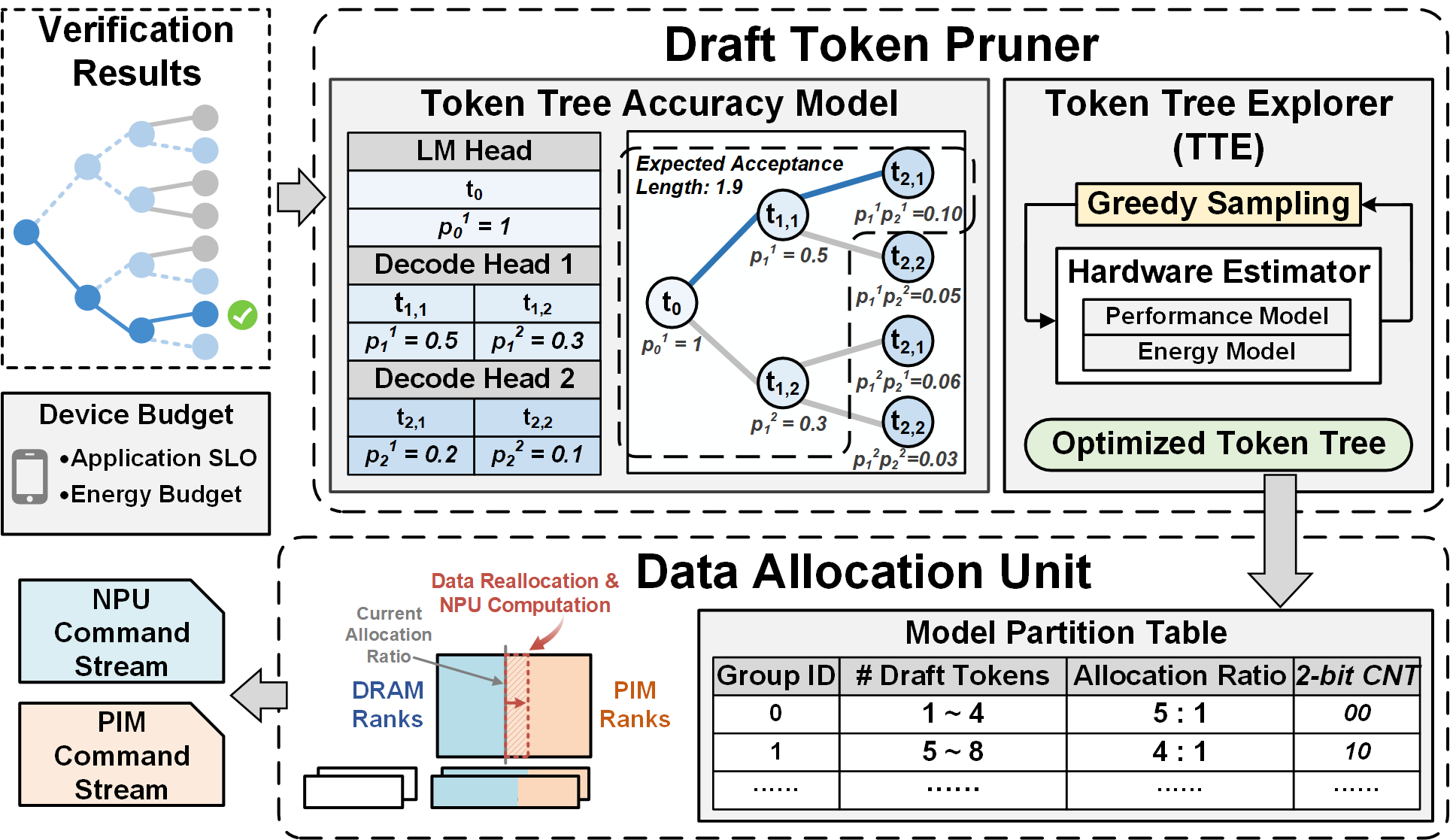}
  \vspace{-5pt}
  \caption{\textit{LP-Spec} workload scheduler consisting of the draft token pruner (DTP) and the data allocation unit (DAU).}
  \label{fig: scheduler framework}
  % \vspace{-10pt}
\end{figure}

\textbf{Token Tree Accuracy Model.} 
We quantify token tree prediction accuracy using expected acceptance lengths for each node in tree $\bm{\mathcal{T}}$ with $N$ nodes. As shown in Fig. \ref{fig: scheduler framework} top, we track the speculation accuracy ${\bm{p}_i}^{k}$ of the $k^{th}$ prediction at $i^{th}$ Decode Head after each decoding step based on previous verification results.
For token $t_{i}$ in a draft sequence, its expected acceptance length is the product sum of the token accuracy in the sequence, i.e., $l_{t_i}= \prod_{i=1}^{m}{\bm{p}_i}^{k_i}$.
For example, $l_{t_{2.1}} = {p_0}^{1}{p_1}^{1}{p_2}^{1}$.
For the entire tree, the expected acceptance length is the sum of all the token nodes.
It is worth mentioning that pruning draft tokens does not incur inference accuracy loss.

\textbf{Hardware Estimator.}
Our hardware estimator combines a performance and energy model to facilitate cost-aware token tree optimization. 
The performance model quantifies NPU-PIM parallel execution constraints, and total execution latency can be formulated as: 
\begin{equation*}
    \bm{T}_{\textit{total}} = min(\bm{T}_{\textit{NPU}}, \bm{T}_{\textit{PIM}})
\end{equation*}

The NPU can be modeled using the roofline model, in which its computational throughput is limited by the off-chip memory bandwidth. 
For PIM devices, the internal bandwidth matches their computation performance.
NPU and PIM execution latency can be modeled as follows:
\begin{equation*}
  \bm{T}_{\textit{NPU}} = \frac{N_{\textit{params, DRAM}}}{\textit{BW}_{\textit{Off-chip}}}, 
  \bm{T}_{\textit{PIM}} = \frac{N_{\textit{params, PIM}}}{\textit{BW}_{\textit{PIM}}} \times  \lceil  \frac{L_{\textit{spec}}}{N_{\textit{ALU}}} \rceil
\end{equation*}
where $N_{\textit{params, DRAM}}$ and $N_{\textit{params, PIM}}$ represent weight/KV-cache volumes used in NPU and PIM computation. $\textit{BW}_{\textit{Off-chip}}$ and $\textit{BW}_{\textit{PIM}}$ are the external and internal bandwidth of LPDDR5-PIM, $N_{\textit{ALU}}$ is the number of ALUs per MPU, which is 4 in our work.
$L_{\textit{spec}}$ represents the speculation length.
The energy model accounts for PIM and NPU computation, and on-chip/off-chip data transfer.

\textbf{Dynamic Token Pruning.}
The DTP operates in a closed-loop manner: Verification results from the previous iteration update accuracy statistics of Decode Heads. These statistics, combined with current device resources and optimization objectives, drive the token tree optimization.
The token tree explorer (TTE) employs a hardware-informed greedy strategy to construct the optimized token tree from root to leaf.
In each sampling step, the TTE adds the node with the highest prediction accuracy to the tree.
Subsequently, the hardware estimator evaluates the performance and energy consumption of speculative inference based on the expected acceptance length of the current token tree. It determines whether to accept the node based on the optimization objectives.
This dynamic pruning process yields an optimized token tree for subsequent decoding steps.

\subsection{Dynamic Workload Scheduling}

To maximize the advantages of the heterogeneous system, we adopt NPU-PIM co-processing, in which FC and attention layers are processed by the NPU and PIM using tensor parallelism. Model weights and KV-cache are partitioned between PIM and DRAM ranks. 
Element-wise and nonlinear functions, such as Softmax and LayerNorm, are executed on the NPU, as implementing these functions in DRAM would incur prohibitive area overhead.
During NPU computation, model parameters are fetched from DRAM ranks, while PIM computation utilizes parameters stored in PIM ranks.

To synchronize the execution time of both NPU and PIM devices, the partition ratio of model parameters across PIM and DRAM ranks needs to correlate with the computation throughput of NPU and PIM devices.
Since PIM computation throughput depends on $L_{\textit{spec}}$, and the token pruning scheme introduces fluctuations in $L_{\textit{spec}}$ across decoding iterations, a fixed model partition ratio can result in sub-optimal performance.
To address this issue, we introduce the data allocation unit (DAU), which dynamically adjusts workload mapping and scheduling between the NPU and PIM based on $L_{\textit{spec}}$.
As shown in Fig. \ref{fig: scheduler framework} bottom, DAU computes the optimal allocation ratio for the next decoding iteration based on the model partition table, which stores the optimal allocation ratios for different $L_{\textit{spec}}$. It then generates the corresponding command stream for synchronized NPU and PIM executions.

To strike a balance between excessive parameter reallocation and improved inference latency, we further implement two optimizations in the DAU. \textit{First}, we use a Group ID to select different partition ratios instead of assigning a unique allocation ratio for each $L_{\textit{spec}}$. Multiple $L_{\textit{spec}}$ values share the same Group ID, and thus the same allocation ratio. \textit{Second}, the DAU employs a 2-bit saturated counter for each $L_{\textit{spec}}$ group to indicate whether the DAU activates. It only activates when $N_{\textit{token}}$ exceeds the threshold consecutively twice.

\begin{figure}[t]
  \centering
  \includegraphics[width=\linewidth]{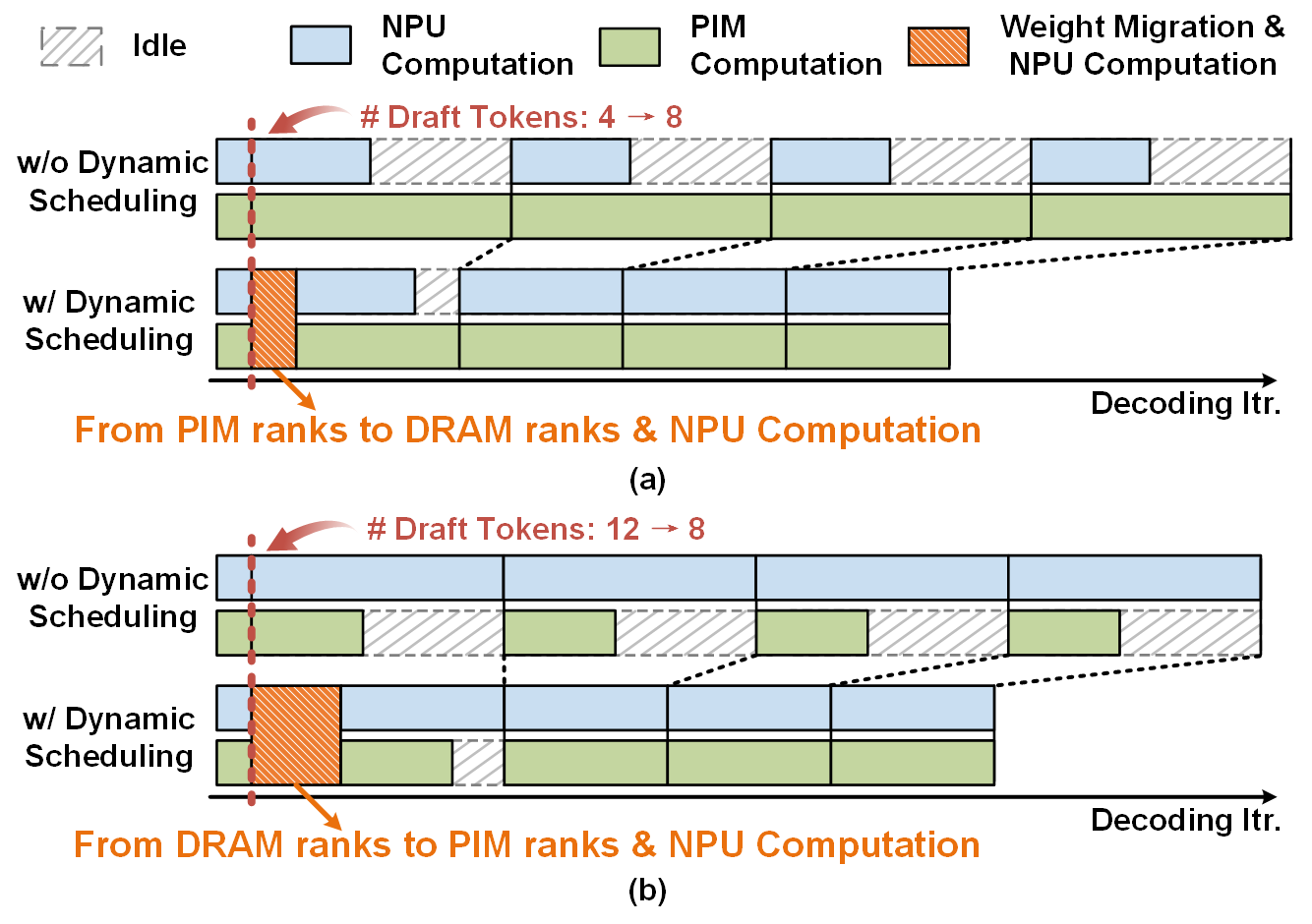}
  \vspace{-10pt}
  \caption{Static workload scheduling of PIM and NPU and our dynamic workload scheduling in two case: (a) draft tokens increases and (b) draft tokens decreases.}
  \label{fig: mapping and scheduling}
  \vspace{-10pt}
\end{figure}

Fig. \ref{fig: mapping and scheduling} illustrates two examples of \textit{LP-Spec} dynamic workload scheduling. When the DAU activates, specific model parameters are reallocated between DRAM and PIM ranks. During this reallocation, the NPU reads the weights from the corresponding DRAM/PIM ranks for computation and incurs copy-write operations to migrate the data to PIM/DRAM ranks via the NMC, effectively overlapping the latency between NPU computation and parameter reallocation.
Compared to parallel execution without dynamic token-aware workload scheduling, both NPU and PIM can become underutilized as $L_{\textit{spec}}$ changes. 
While the parallel execution during the decoding iteration that involves data reallocation may be imbalanced, the saturated counter ensures that $L_{\textit{spec}}$ does not fluctuate dramatically in subsequent iterations, thereby maintaining synchronization between PIM and NPU.

    \begin{figure*}[t]
  \centering
  \includegraphics[width=\linewidth]{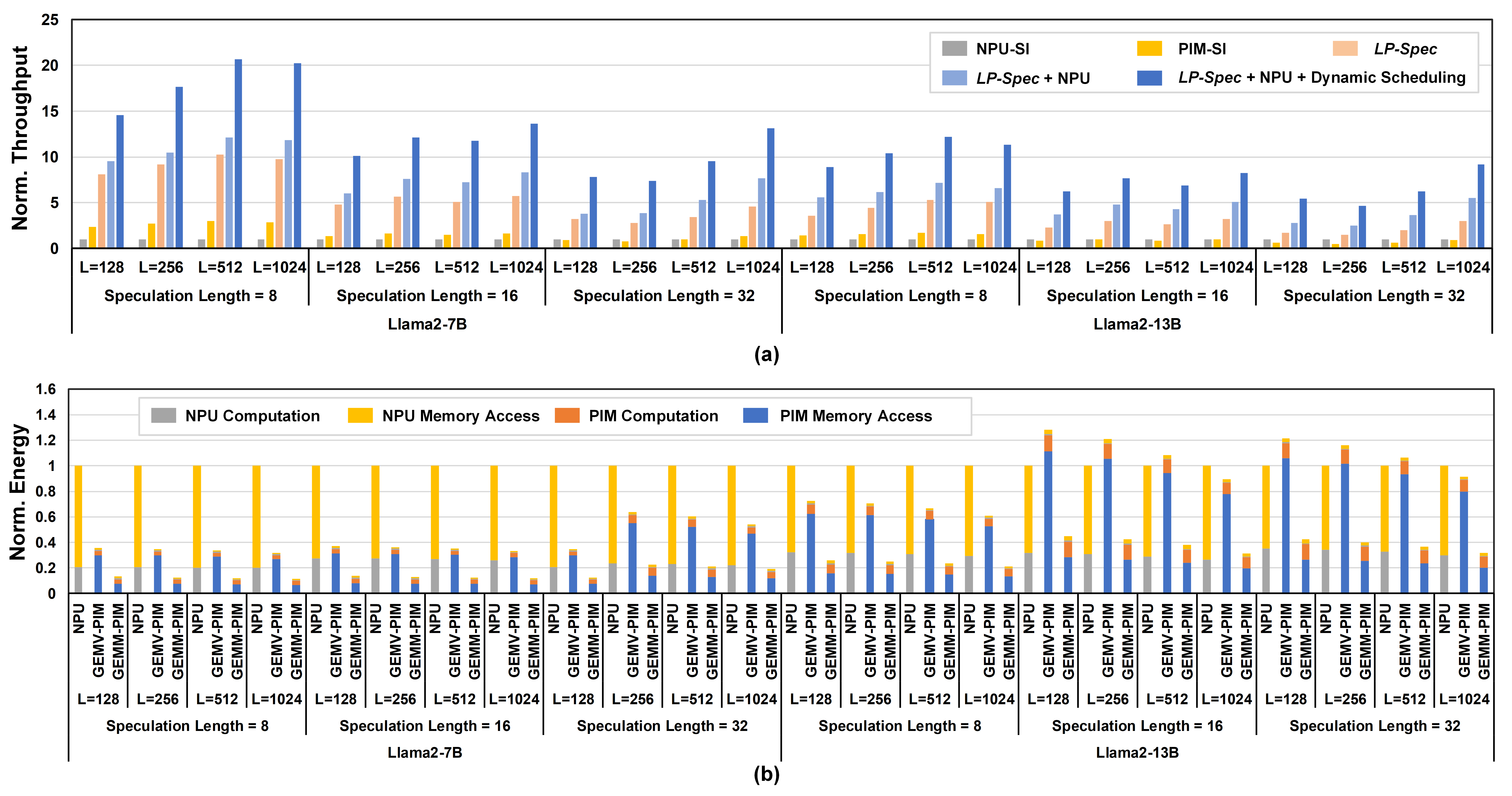}
  \vspace{-10pt}
  \caption{Throughput and energy efficiency improvements compared with baselines.}
  \label{fig: perf and energy comparison}
  \vspace{-5pt}
\end{figure*}

\section{Evaluations}
\subsection{Experimental Setup}
In this work, we implement our \textit{LP-Spec} architecture based on established DRAM vendor designs \cite{kim_samsung_2023}, with specifications listed in Table \ref{tab: hardware specification}. We design \textit{LP-Spec} using $\times$64 LPDDR5 to represent typical mobile device scenarios, e.g., laptops and smartphones.
Compared with prior PIM designs\cite{kim_samsung_2023}, we enhance the performance by 4$\times$ to 409.6 GOPS for each die.
We develop an in-house simulator by integrating and extending the Samsung PIM simulator \cite{samsung_pim} and LLMCompass \cite{llmcompass} for \textit{LP-Spec} system evaluation. Energy consumption data are obtained from prior works \cite{park_attacc_2024, li_specpim_2024, mcdram2}.
For area evaluation, we use Cadence Genus to synthesize the \textit{LP-Spec} bank-level MPU with a 7\textit{nm} process \cite{clark_asap7_2016} at 200 MHz given $t_{\textit{CCD}}$ (5\textit{ns}). The area overhead for compute units and buffers on the DRAM die was scaled to 1\textit{z-nm} DRAM process, considering that the DRAM process is 10$\times$ less dense than the logic process \cite{upmem}.
The total area overhead is 10.31 \textit{mm}$^{2}$ per DRAM die, which translates to 16.92\% of a 60.92 \textit{mm}$^{2}$ 16 Gb LPDDR5 die \cite{area_lpddr5}.
Based on chip microphotograph analysis of DRAM and PIM dies \cite{lee_hardware_2021, area_lpddr5}, this design reduces memory density by 26.5\% compared to Samsung LPDDR5-PIM dies.

\begin{table}[t]
  \caption{\textbf{Hardware Specification for \textit{LP-Spec}}}
  \label{tab: hardware specification}
  % \vspace{-8pt}
  \renewcommand{\arraystretch}{1.2}
  \resizebox{1\linewidth}{!}{
  \begin{tabular}{cccc}
    \hline
    \multicolumn{4}{c}{\textbf{NPU Configuration}}                                                                                \\
    \hline
    \multicolumn{1}{c|}{Matrix Unit}                                                                                             & \multicolumn{1}{c|}{32.8 TFLOPS} & \multicolumn{1}{c|}{Vector Unit}           & 8.2 TFLOPS          \\
    \multicolumn{1}{c|}{Compute Core / Chip}                                                                                     & \multicolumn{1}{c|}{16}          & \multicolumn{1}{c|}{Local Buffer Capacity} & 256 KB              \\
    \multicolumn{1}{c|}{Operation Frequency}                                                                                     & \multicolumn{1}{c|}{1 GHz}       & \multicolumn{1}{c|}{Scratchpad Capacity}   & 8 MB                \\
    \hline
    \multicolumn{4}{c}{\textbf{LP-Spec PIM Configuration}}                                                                        \\
    \hline
    % \multicolumn{1}{c|}{Die Capacity} & \multicolumn{1}{c|}{1 GB} & \multicolumn{1}{c|}{\# Dies} & 16 \\
    \multicolumn{1}{c|}{\# MPU}                                                                                       & \multicolumn{1}{c|}{8}          & \multicolumn{1}{c|}{\# PIM Units / MPU}     & 8 \\
    \multicolumn{1}{c|}{MAC Operation Frequency}                                                                                 & \multicolumn{1}{c|}{200 MHz}     & \multicolumn{1}{c|}{Performance}      & 409.6 GOPS@INT8         \\
    \multicolumn{1}{c|}{On-chip Bandwidth}                                                                                       & 51.2 TB/s                        & \multicolumn{1}{|c|}{Capacity}   & 1 GB           \\
        
    \multicolumn{1}{c|}{\# Die / Rank}                                                                                       & 4                        & \multicolumn{1}{|c|}{Capacity / Rank}   & 4 GB           \\
    \hline
    \multicolumn{4}{c}{\textbf{Samsung LPDDR5-PIM Configuration\cite{kim_samsung_2023} }}                                         \\
    \hline
    \multicolumn{1}{c|}{\# PIM Unit / Die}                                                                                       & \multicolumn{1}{c|}{8}          & \multicolumn{1}{c|}{Performance}      & 102.4 GOPS@INT8         \\
    \multicolumn{1}{c|}{On-chip Bandwidth / Die}                                                                                       & 51.2 GB/s                        & \multicolumn{1}{|c|}{Capacity / Die}   & 1 GB           \\
    
    \multicolumn{1}{c|}{\# Die / Rank}                                                                                       & 4                        & \multicolumn{1}{|c|}{Capacity / Rank}   & 4 GB           \\
    \hline

    \multicolumn{4}{c}{\textbf{LPDDR5 DRAM Configuration}}                                         \\
    \hline
    \multicolumn{1}{c|}{Off-chip Bandwidth}                                                                                       & 51.2 GB/s                        & \multicolumn{1}{|c|}{Capacity / Die}   & 1 GB           \\
    
    \multicolumn{1}{c|}{\# Die / Rank}                                                                                       & 4                        & \multicolumn{1}{|c|}{Capacity / Rank}   & 4 GB           \\
    \hline
    \multicolumn{4}{c}{\textbf{PIM / DRAM Timing Parameter}}                                                                             \\
    \hline
    \multicolumn{4}{c}{$t_{RP}$=15, $t_{RCD}$=15, $t_{RAS}$=34, $t_{RRD}$=4, $t_{WR}$=28, $t_{RC}$=30, $t_{CCD}$=4, $t_{FAW}$=16} \\
    \hline
  \end{tabular}
  }
  % \vspace{-15pt}
\end{table}

In our work, we evaluate our \textit{LP-Spec} against two baselines: (1) Speculative inference on NPU (\textit{NPU-SI}), (2) Speculative inference on GEMV-PIM (\textit{PIM-SI}). Prefill stage of LLM inference and nonlinear functions are executed on the NPU.
The baseline mobile NPU is modeled based on commercial products \cite{huang_5g_2023} with a total performance of 41 TFLOPS, as detailed in Table \ref{tab: hardware specification}, while Samsung LPDDR5-PIM is used as the GEMV-PIM baseline.
For workload evaluations, we adopt the Medusa \cite{cai_medusa_2024} framework with Llama2-7B and Llama2-13B \cite{zheng_judging_2023} using INT8 to profile speculative inference on real-world LLM datasets, e.g., Alpaca \cite{taori_stanford_2023}.
Our \textit{LP-Spec} can also be implemented using other speculative inference strategies with minimal effort, e.g., DLM-based speculative inference.

\subsection{Design Benefits of \textit{LP-Spec}}
We compare the end-to-end performance of \textit{LP-Spec} against the above baselines with different ($L_{in}$, $L_{out}$) and speculation lengths for two LLMs, in which the DRAM memory configuration is set to 3 PIM ranks and 1 DRAM rank with a total capacity of 16 GB. 
As shown in Fig. \ref{fig: perf and energy comparison}(a), \textit{LP-Spec} outperforms the conventional system, achieving on average 4.59$\times$ and 3.25$\times$ over \textit{NPU-SI} and \textit{PIM-SI} baselines, respectively.
Compared with the NPU baseline, \textit{LP-Spec} can utilize the high memory bandwidth of PIM while the NPU baseline is limited by the off-chip memory bandwidth.
Although \textit{PIM-SI} enjoys the same bandwidth benefits as \textit{LP-Spec}, its computational limitations degrade speculative inference performance, causing its advantage over the NPU baseline to diminish as speculative length increases, which is even worse than \textit{NPU-SI} at the speculative length of 32.
However, \textit{LP-Spec} combines high internal memory bandwidth with enhanced compute resources, making it particularly effective for speculative inference across different speculation lengths.
Besides, NPU-PIM co-processing and \textit{LP-Spec} scheduling further improve the end-to-end performance of \textit{LP-Spec}.
Compared to the naive \textit{LP-Spec}, NPU-PIM co-processing achieves an end-to-end performance improvement up to 1.47$\times$ in average.
\textit{LP-Spec} scheduling demonstrates an additional improvement up to 2.49$\times$ in average.
When combining them, \textit{LP-Spec} with NPU-PIM co-processing and \textit{LP-Spec} scheduling achieves performance improvement of up to 13.21$\times$ (8.47$\times$) and 7.91$\times$ (8.33$\times$) over \textit{NPU-SI} (\textit{PIM-SI}) baselines for Llama-2 7B and Llama-2 13B.
For the following analysis, \textit{LP-Spec} refers to the \textit{LP-Spec} architecture with these two optimizations.

Fig. \ref{fig: perf and energy comparison}(b) shows the energy consumption of \textit{NPU-SI}, \textit{PIM-SI}, and \textit{LP-Spec}.
Compared with \textit{NPU-SI}, \textit{LP-Spec} achieves on average 7.56$\times$ improvement due to less off-chip data transfer.
Against \textit{PIM-SI}, \textit{LP-Spec} delivers on up to 2.85$\times$ improvement as our optimized PIM architecture can capture more data reuse opportunities, thereby minimizing the DRAM internal memory accesses.
As the speculative length increases, \textit{PIM-SI} advantage over \textit{NPU-SI} degrades.
This is because \textit{NPU-SI} can fully reuse weight data to reduce off-chip memory access energy, while \textit{PIM-SI} presents limited reuse opportunities.
In contrast, \textit{LP-Spec} replaces vector units in prior PIM designs with MPUs, improving weight reuse rates and reducing internal DRAM memory access overhead.
Meanwhile, the DTP in the \textit{LP-Spec} scheduler largely reduces energy on verifying rejected tokens.

\begin{table}[]
  \centering
  \caption{Comparison with Other Acceleration Hardware}
  \label{table: compare with other works}
  \renewcommand{\arraystretch}{1.1}
  \begin{tabular}{cccc}
    \toprule                                & \textbf{AttAcc} \cite{park_attacc_2024} & \textbf{RTX 3090} & \textit{\textbf{LP-Spec}} \\
    \midrule \textbf{Target Scenario}       & Cloud                                   & Mobile            & Mobile                    \\
    \textbf{Model}                          & GPT-3 175B                              & Llama2-7B         & Llama2-7B                 \\
    \textbf{Throughput (token/s)}           & 33.3                                    & 44.3              & \textbf{73.4}             \\
    \textbf{Energy Efficiency (token/J)}    & 5.6                                     & 0.13              & \textbf{32.6}             \\
    \textbf{EDP (s$\cdot$mJ ) $\downarrow$} & 5.36                                    & 173.6             & \textbf{0.418}            \\
    % & 186.5   & 5.8     & \textbf{1015.6}   \\
    \bottomrule                              % \vspace{-25pt}
  \end{tabular}
\end{table}

\subsection{Discussions}

Prior designs \cite{park_attacc_2024,li_specpim_2024,heo_neupims_2024} have mainly focused on the PIM-GPU system for cloud services.
In these designs, compute-intensive FC layers are offloaded to GPUs, which are equipped with high-bandwidth HBM or GDDR.
However, this method cannot be adapted for mobile devices due to low memory bandwidth.
Consequently, we develop \textit{LP-Spec} with hardware-dataflow co-optimization architecture to address the above challenges for mobile devices.

As summarized in Table \ref{table: compare with other works}, \textit{LP-Spec} demonstrates a competitive throughput and a superior energy efficiency of token/J for Llama2-7B inference, which achieves 12.83$\times$ and 415.31$\times$ better EDP than the autoregressive decoding on AttAcc \cite{park_attacc_2024} and RTX 3090. Overall, \textit{LP-Spec} is an efficient approach for deploying LLM speculative inference on mobile devices, which can also be extended to other GEMM-dominated workloads, e.g., vision Transformers.

    \section{Conclusion}
In this paper, we analyzed the challenges of deploying speculative inference on mobile devices, which can not efficiently process GEMM operations on existing hardware.
% Thus, we present \textit{LP-Spec}, an efficient LPDDR5-based PIM-NPU architecture with 8$\times$ performance-enhanced PIM architecture and hardware-aware runtime token pruning strategy.
We present \textit{LP-Spec}, a NPU-PIM heterogeneous architecture with GEMM-enhanced LPDDR5-PIM and a dedicated near-data memory controller to accelerate speculative inference and improve the energy efficiency.
We also adopt draft token pruning to prune the redundant tokens to improve the energy efficiency and performance.
Besides, the dynamic scheduling is proposed to improve hardware utilization of NPU-PIM systems.
% reducing the inference latency.
% 8$\times$ performance-enhanced PIM architecture and hardware-aware runtime token pruning strategy.
Evaluations demonstrate that \textit{LP-Spec} achieves up to 13.21$\times$, 7.91$\times$ performance and 7.56$\times$, 2.85$\times$ energy efficiency gains over the mobile NPU and PIM baselines.
Compared with other existing works, \textit{LP-Spec} exhibits 12.83$\times$ and 415.3$\times$ better EDP than AttAcc PIM and 3090 GPU.

% In this work, we present \textit{LP-Spec}, a GEMM-enhanced PIM architecture with in-memory PE array to facilitate both GEMM and GEMV computation for edge LLM inference. 
% A locality-aware workload partitioning and execution flow is further developed to minimize data communication.
% Furthermore, a hardware-aware token pruning strategy is developed to minimize energy consumption on resource constrained mobile devices.
% Compared to end-to-end inference on both mobile NPUs and heterogeneous systems with NPU and GEMV-accelerated PIM device, \textit{LP-Spec} demonstrates promising performance gains with notable energy savings.

\section*{Acknowledgement}
This work was supported in part by NSFC No. 92464202.

    \bibliographystyle{ieeetr}
    \bibliography{reference}
\end{document}